\documentclass[twoside,12pt]{article}
\usepackage{epsfig}
\newcommand{\be}{\begin{equation}}
\newcommand{\ee}{\end{equation}}
\newcommand{\bea}{\begin{eqnarray}}
\newcommand{\eea}{\end{eqnarray}}
\newcommand{\eq}[1]{(\ref{#1})}
\newcommand{\ubar}{\overline u}
\newcommand{\dbar}{\overline d}
\newcommand{\psibar}{\overline \psi}
\newcommand{\E}{\mathrm{e}}
\newcommand{\I}{\mathrm{i}}
\newcommand{\srm}[1]{\textrm{\scriptsize #1}}
\newcommand{\FD}{\;.}
\newcommand{\FC}{\;,}
\newcommand{\Dslash}{\ensuremath{D\hspace{-8pt}/}}
\newcommand{\ot}{\ensuremath{\frac{1}{2}}}

\begin{document}

\title{\vspace{1cm} The Hadron Spectrum from Lattice QCD\thanks{Notes based on a lecture
at the Int. School of Nuclear Physics, 29th Course, 16-24. Sept. 2007, Erice/Sicily, 
``Quarks in Hadrons and Nuclei''}}
\author{C.\ B.\ Lang\\
\\
Inst. f. Physik, FB Theoretische Physik,\\ 
Universit\"at Graz, A-8010 Graz, Austria}
\maketitle
\begin{abstract} 
Determining the hadron spectrum and hadron properties beyond the ground states  is
a challenge in lattice QCD. Most of these results have been in the quenched
approximation but now we are entering the dynamical era. I review some of the
ideas and methods of the lattice approach, concentrating on a few examples and
on results obtained for Chirally Improved (CI) fermions. 
\end{abstract}

\section{Introduction}
\subsection{\it Motivation}
This lecture reviews some methods to compute the masses of hadrons in
lattice QCD. The emphasis lies on the low lying excited states. In the spirit
of the school I try to combine an introduction to the field with a
presentation of some recent work. Most of the results for masses presented here
are based on work within the Bern-Graz-Regensburg (BGR) collaboration. As a
motivation let me start with some results obtained for the so-called quenched
situation, where the fermion loops of the vacuum are disregarded.

\begin{figure}[tb]
\begin{center}
\epsfig{file=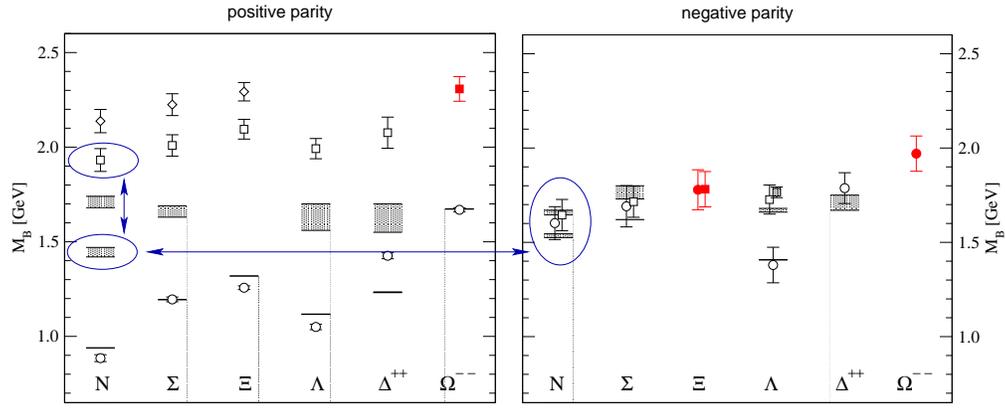,scale=0.58}
\caption{The chiral extrapolations of the positive parity (left) and
negative parity (right) low lying baryons \cite{BuGaGl06b}.
\label{fig:quebarspec}}
\end{center}
\end{figure}

In that approximation the quarks are valence quarks and it is comparatively
simple to work at several quark masses. Since one cannot (yet) compute at small
enough quark masses, one extrapolates the values obtained from higher masses
(corresponding to pion masses around and above 300 MeV) down to the fictitious
chiral limit, where the current quark masses of the up and the down quarks
vanish. Since this is close to the physical case, it provides a useful standard
for comparison of different computations. Fig.\ \ref{fig:quebarspec} gives the
results for the low lying baryons (in the situation $m_u=m_d=0$ and $m_s$ such
that the kaon mass has its experimental value). 

There are some observations to be made:
\begin{itemize}
\item Ground state and first excitations are clearly seen. The negative parity
states are close to their experimental values. Four states (positive parity:
$\Omega$ excitation near 2.3 GeV, negative parity: two $\Xi$ states near 1.78
GeV and a $\Omega$ ground state near 1.97 MeV) are seen, which are not
established in the  Particle Data Group tables. Both negative parity states
$N(1535)$ and $N(1650)$ are seen.
\item The excitations in the positive parity sector are clearly too high. In
particular the Roper state $N(1440)$ comes out 30\% too heavy. Since the standard
level ordering is seen for the heavy quark region, one expects a level crossing of
the $N(1535)$ and the Roper towards physical quark masses. This is not observed in
these calculations.
\end{itemize}
The usual suspects are finite volume effects and the missing dynamical fermions.

\begin{figure}[tb]
\begin{center}
\epsfig{file=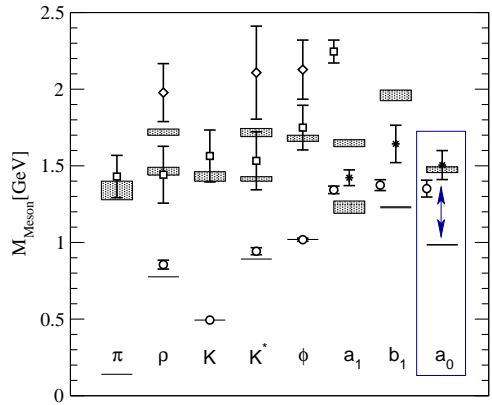,scale=0.65}
\caption{The chiral extrapolations of the 
low lying mesons \cite{BuGaGl06}.\label{fig:quemesspec}}
\end{center}
\end{figure}

Fig. \ref{fig:quemesspec} gives the meson results for that quenched study
\cite{BuGaGl06}. Here we find that the ground state of the isovector scalar
meson $a_0$ (emphasized by the surrounding box in the figure)
is either not seen or comes much too high, i.e., where the first
excitation $a_0(1450)$ is measured. Again it would be important to find 
out how fully dynamical sea-quarks would affect this picture.

Let us summarize the challenges and questions:
\begin{itemize}
\item What effects have dynamical quarks on the excitations, in particular
towards smaller, physical quark masses?
\item What is the role of chiral symmetry (and its breaking)? Is there a
``symmetry restoration'' \cite{Gl07} towards higher excitations?
\item How can we improve the analysis method and get even higher excitations?
How to get properties like coupling constants and decay constants for the
excited states?
\end{itemize}

\subsection{\it Lattice QCD}

Let me start with a ``Declaration of QCD'': We assume that QCD is {\it the}
quantum field theory of quarks and gluons, defined by a Lagrangian and
action of the form
\bea
L&=&\frac{1}{2\,g^2}\, \mathrm{tr} F_{\mu\nu} F_{\mu\nu} +\sum_f
\overline\psi_f \;(\Dslash + m_f) \;\psi_f\;,\\
S &=&\int d^4x \;L\;.\nonumber
\eea
We have formulated the theory in Euclidean space-time for convenience. The sum
is over the quark flavors, $\Dslash$ denotes the usual Dirac operator. We
furthermore assume that this theory can be solved from first principles, with a
minimal number of experimental input parameters (the bare quark masses and a
scale fixing parameter) and that all hadron properties should be computable that
way.

\begin{figure}[b]
\begin{center}
\epsfig{file=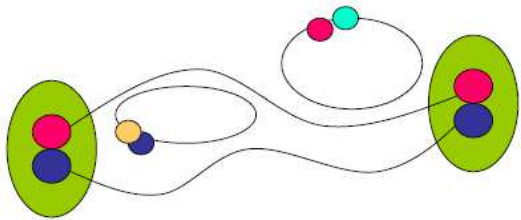,scale=0.5}
\caption{Schematic propagation of a meson. The closed quark-loops are neglected
in the so-called quenched approximation.\label{fig:pathint}}
\end{center}
\end{figure}

Quantization of such a field theory may be done with a path (functional)
integration. A  two-point function describing the propagation of a nucleon
would be evaluated through
\be\label{eq:corrfunction}
C_N(t)=\langle\overline N_t N_0\rangle 
\propto \int[\mathcal{D}A \,\mathcal{D}\overline\psi\,
\mathcal{D}\psi]\;\mathrm{e}^{-S(A,\overline\psi,\psi)} \;\overline N_t\,N_0
\sim \exp(-E_N\,t)\;.
\ee
Assuming that the nucleon is at rest, the asymptotic exponential behavior is
determined by its mass. In Fig.\ \ref{fig:pathint} such  a propagation of a meson is
shown schematically. In the path integral fermions are represented by
Grassmann-variables and can be integrated explicitely. The result is the
determinant of the Dirac operator, which acts as a weight in the path integral
over the gauge fields. Neglecting the determinant amounts to neglecting the
vacuum quark loops, the so-called quenched approximation.

Ken Wilson suggested more than 30 years ago \cite{Wi74} to formulate QCD on a
Euclidean space time lattice. This way the functional integral is 
high-dimensional but well defined. The quark fields $\psi$ become Grassmann
variables defined on the lattice sites and the color gauge fields are
represented by SU(3)-matrices $U_\mu(x)$ living on the links connecting the
sites. The Dirac operator becomes just a (very, very large) matrix. This
formulation opened the way to non-perturbative path integration with the  help
of computers. Obviously it is an approximation and one has to study whether
the approximation makes sense and leads to stable results in the continuum
limit, as discussed below. Lattice QCD provides a gauge invariant definition of
QCD as a quantum field theory. As formulated it is an approximation, which may
be improved by various techniques, but it  is QCD and not a model of
QCD-phenomenology.

A few years after Wilson's suggestion actual calculations were done for
Yang-Mills theory, starting with Mike Creutz' seminal suggestion \cite{Cr80}
how to use Monte Carlo methods to compute the  path integral for the gauge
fields. Meanwhile lattice field theorists have entered prominently the crowd of
high demand and high performance computer users. Quenched calculations have
become well controlled. There the gluon interactions are
covered completely non-perturbatively. Valence quarks on the gluonic background
are done in various contexts and many of the results have been -- given the
quenched approximation -- in surprising agreement with experiments. The issue
nowadays is to include fermions fully dynamically, i.e., including the fermion
determinant in the path integral weight.

In QCD there are $n_f+1$ parameters, corresponding to the scale and the
$n_f$ current quark masses. In the lattice formulation these are dimensionless
parameters like the gauge coupling and the bare quark masses. The calculation
produces n-point functions, and the first results have been for propagators of
QCD bound states like the pion or the nucleon. The exponential decay
\eq{eq:corrfunction} gives the dimensionless exponent $M\,t = (a\, M) (t/a)$,
where $a$ denotes the lattice spacing  in physical units.

We thus get dimensionless masses $a M$ from this analysis and have to set the
scale $a$. The lattice spacing changes with the bare parameters of the
simulations and these have to be tuned in order to approach the continuum limit phase
boundary, which is where $a(g,\,m)\to 0$ ($m$ symbolizes the set of all bare
quark masses). Optimally one would follow trajectories in the parameter space
where mass ratios are fixed to their physical values. Due to the enormous
computational cost this is not always possible and one relies on patchwork
combining results for different domains of the parameters.

Given the physical mass $M$ and from the measurement the dimensionless lattice
mass $a M$, one can determine the lattice spacing for the simulation set of
parameters. Using $n_f+1$ physical numbers (like, e.g., bound state
masses or decay constants) one can, in principle, fix all unknown quantities
and therefrom compute all physics contained: other masses, decay constants, matrix
elements,  scattering amplitudes and so on.

Practical calculations are restricted to finite sets of numbers and finite
computer resources. In that sense the lattice simulations are approximations.
We are interested in understanding three limits in our calculations.

\begin{figure}[b]
\begin{center}
\epsfig{file=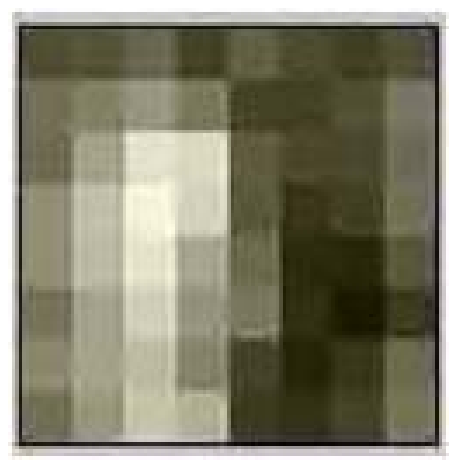,scale=0.5}
\epsfig{file=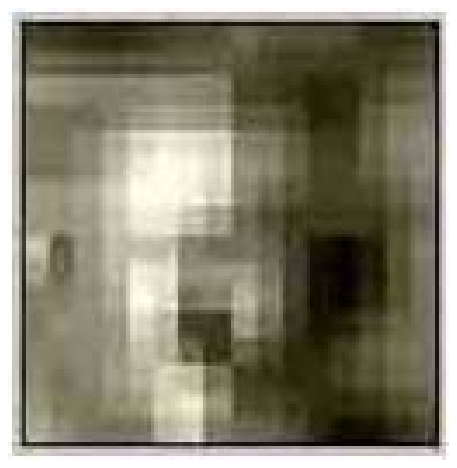,scale=0.5}
\epsfig{file=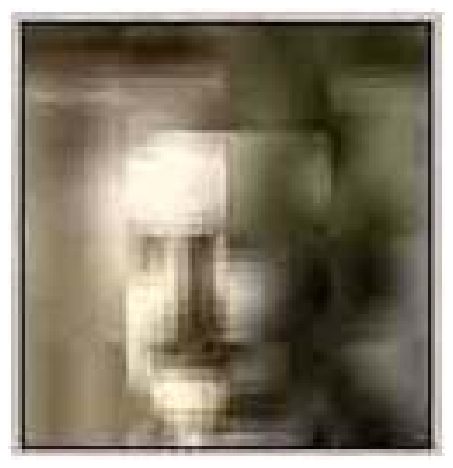,scale=0.5}
\epsfig{file=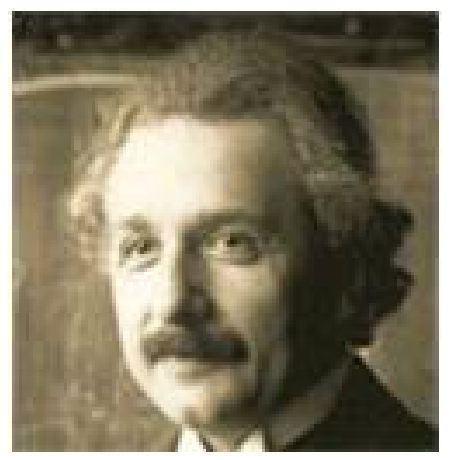,scale=0.5}
\caption{The same physical image represented on lattices of linear extent
8, 16, 32, and 128 corresponding to lattice spacings of 4 cm, 2 cm, 1cm, 
and 1/4 cm.
\label{fig:contlim}}
\end{center}
\end{figure}

\begin{description}
\item[Continuum limit $a(g,\,m)\to 0$:] This is obtained for the bare gauge
coupling $g\to 0$ and the mass parameters adjusted such that mass ratios remain
constant. Different lattice actions correspond to different quality of
approximations. In a perturbative expansion the Wilson gauge action has
corrections $\mathcal{O}(a^2)$ and the Wilson fermion actions $\mathcal{O}(a)$.
There are various attempts to improve that behavior, mostly based on either
perturbative concepts (the Symanzik improvement program) or non-perturbative
concepts (like real space renormalization transformations). There are now
improved fermion actions which have corrections of only $\mathcal{O}(a^2)$. One
hopes to obtain results closer to the continuum situation if these
corrections are small. Typical lattice spacing values in full QCD simulations
lie between $0.07$ and $ 0.2$ fm.
\item[Thermodynamical limit $L\to\infty$:] Assuming the lattice has  $L^4$
sites, its physical extension is $a\,L$. Ideally one wants to keep that
physical size reasonable large (several fm) and constant when $a\to 0$.
Therefore $L$ has to grow correspondingly when reducing the  lattice spacing
(cf. Fig.\ \ref{fig:contlim}).
\item[Chiral limit $m\to 0$ or physical quark mass limit $m\to m_0$:] It is substantially
harder to simulate at smaller quark masses. This is related to numerical
problems like matrix inversion and to the effect of increasing correlation
lengths (smaller pion mass). For that reason one has so far worked at
relatively large bare quark masses corresponding to pion masses around 400 MeV
and higher. On the other hand, real life occurs at small quark masses and a
pion of 140 MeV. This is close to the chiral limit where quarks and pions are
massless due to spontaneous chiral symmetry breaking.

In that limit we have some understanding of the effective theory, notably chiral
perturbation theory (ChPT) \cite{We79GaLe84}. This helps to analyze the
approach from higher  towards smaller quark masses. Present day results indicate
that one has to have pion masses below 300 MeV to clearly identify terms of the
effective theory.
\end{description}

\section{Fermions and chiral symmetry}

\subsection{\it The Ginsparg-Wilson condition}

In continuum the massless Dirac operator anti-commutes with $\gamma_5$,
\be\label{eq:chiralcont}
\Dslash\,\gamma_5+\gamma_5\,\Dslash=0\FD
\ee
This leads to a decoupling of left-handed and right-handed modes. If the light
quarks are  massless, the chiral symmetry $SU(2)_L \times SU(2)_R$ is broken
spontaneously by the strong interaction and there is a massless Goldstone
triplet, the pion. In reality the light quarks have small, but non-vanishing
masses leading to an additional explicit breaking of the symmetry giving rise to
the light pions. This behavior may be described by ChPT.

In order to show that there is spontaneous symmetry breaking, a theory should
allow for chiral symmetry for vanishing quark masses. Otherwise we cannot be
sure, that the mechanism is as expected. Unfortunately the lattice formulation
does not allow for a chirally symmetric formulation like in the continuum. There is a
fundamental theorem \cite{NiNi} stating that under quite
general conditions chiral symmetry on the lattice cannot be realized without
invoking additional fermions (so-called doublers). The original formulation of
Wilson pairs left-handed and right-handed fermions.

In a seminal paper Ginsparg and Wilson \cite{GiWi82} suggested a condition
which formulates chiral symmetry on the lattice in a modified form. Based on
renormalization group arguments they proposed to replace \eq{eq:chiralcont} by
another equation, which in a simple version may be written
\be\label{eq:chirallatt}
D\,\gamma_5+\gamma_5\,D= a\,D\,\gamma_5\,D\FD
\ee
The right-hand side vanishes in the continuum limit and full continuum chiral
symmetry is recovered. This suggestion was considered infeasible in practice
and the paper was essentially ignored until the late 1990ies, when it was realized
that there are indeed possible realizations \cite{HaNi94HaHaNi05,Ne}
of that concept.  Eventually a lattice version of the chiral rotations was
formulated \cite{Lu98} which leaves the massless lattice action, if obeying the
Ginsparg-Wilson condition (GWC), invariant.

\subsection{\it Fermion species}

What fermions are on the market and what are their advantages or disadvantages?
There are two main groups: GW-type and non-GW-type. In the first group the most
prominent representatives are 
\begin{itemize}
\item Wilson improved: The original Wilson fermion action which  does have
doubler modes with masses $\mathcal{O}(1/a)$, improved by an extra term which
has the correct dimension to correct for chiral symmetry. Its coefficient has
to be adjusted non-perturbatively which corrects the additively renormalized
fermion mass. One of the technical problems is the occurrence of spurious zero
modes of the Dirac operator for smaller (but non-zero) quark masses. The
spurious modes turn up randomly for the sampled gauge configurations. This
prevents computations at small quark masses unless one goes to very large
lattices and small lattice spacing, where the situation improves.
\item Staggered fermions: The fermionic degrees of freedom are distributed over
hypercubes and a remnant chiral symmetry is sustained. Since there is no
additive fermion mass renormalization, this formulation is convenient to
simulate. There are, however, still too many fermions (a factor of 4 instead of
16 for the Wilson fermions) and these ``tastes of fermions'' confuse the identification of
the hadron states. It is presently disputed, whether the theory has a correct
continuum limit.
\item Twisted mass fermions: An extra term in the Wilson action changes  the
eigenvalue spectrum of the Dirac operator such that there are no spurious
modes. The theory breaks parity and flavor symmetry; this is recovered
in the continuum limit, though.
\end{itemize}

The most prominent representative in the group of GW-type actions is the  {\bf
overlap Dirac operator}  \cite{Ne}. The massless operator is an explicit
construction,
\be
D_\srm{ov} = \frac{1}{a}\left[1+\gamma_5\,\mathrm{sign}(\gamma_5\, A)\right]\FC
\ee
where $A$ is a kernel operator like, e.g., the usual Wilson operator, evaluated
at negative quark masses. The technical problem is to compute the sign function
of the huge matrix $(\gamma_5\, A)$. So far this operator has been used mainly
for quenched calculations, i.e., for the valence quark propagators. Is is
technically very demanding to implement it for dynamical fermions and attempts
in this direction are under progress by various groups. With the massless
overlap operator one may have single left- or right handed zero modes.

The  {\bf domain wall action} \cite{Ka92,FuSh95} uses an extra 5th dimension in
order to separate left-handed from right-handed modes and approaches the
overlap operator in the limit of infinite extension in the 5th direction.

A perfect operator would be one that lies on a renormalized trajectory.  An
approximation to that is the {\bf fixed point operator} \cite{HaNi94HaHaNi05},
which lies close to such a trajectory and also has good, but not perfect,
chiral properties.

Another approximate GW-type operator is the so-called   {\bf chirally improved
operator} (CI-fermions) \cite{Ga01aGaHiLa00}, which has been constructed from
a systematic expansion in terms of nearest-neighbor, next-to-nearest-neighbor
etc. terms, plugged into the GWC and solved algebraically. Since the expansion
has to be truncated, the GWC is obeyed only approximately, like in the case of
the fixed point or the  domain wall action.

In the BGR collaboration we have been studying both, the fixed point and the
CI operators \cite{BGR04}. Both show good chiral behavior
and allow to approach smaller pion masses than is possible with the improved
Wilson action for same size lattices. The actions have many more terms and are
an order of magnitude more expensive to simulate than
the simpler non-GW-type actions. However, they are still about
an order of magnitude cheaper than expected for the overlap action. 
For the rest of this presentation I will restrict myself to showing results from my
collaboration based on CI fermions.

\section{Excited states}

\subsection{\it How to obtain particle masses}

Masses are measured through real space propagators of the form
\be
C_X(\vec p,t)=\langle
X(\vec p,\,t)\, \overline X(\vec p,\,0)
\rangle
\quad 
\textrm{with}
\quad
X(\vec p,\,t)=\sum_{\vec x}\, \hat X(\vec x,\,t)\,\E^{-\I\, \vec x\cdot\vec p}\;.
\ee
The hadron interpolators $X$ are chosen with specified Lorentz-, Dirac-,
flavor-, and color symmetries. We have projected the real space operators to
a fixed 3-momentum $\vec p$. The propagation functions will vanish if conservation laws
due to these symmetries are violated, e.g.,  if one combines an operator with
the quantum numbers of a pion as source with that of a rho-meson as a sink.
Simple meson field operator will have a form like $\dbar(x) \gamma_5 u(x))$
for the $\pi^+$ but can be more complicated involving also non-local but gauge
invariant terms like $\dbar(x) \gamma_5 U(x,y)u(y)$ where $U$ denotes
a product of gauge field link variables connecting $x$ with $y$.

When integrating over the Grassmann variables, the correlation function is
expressed in terms of matrix elements of the inverted Dirac operator, i.e., the
quark propagator. Eq. \eq{eq:corrfunction} becomes an integral over the gauge
field variables with $\mathrm{det}(D + m_f)$ as weight factor and products
of quark propagator matrix elements  $(D + m_f)^{-1}_{xy}$ in the
integrand,
\be
C_\pi(x \to y) 
\propto \int[\mathcal{D}U ]\;\E^{-S_\srm{gauge}(U)} \left(\prod_f 
\mathrm{det}(D + m_f)\right)
\mathrm{tr}\left( (D + m_u)^{-1}_{xy}\, \gamma_5\, (D + m_d)^{-1}_{yx} 
\,\gamma_5\right)\FD
\ee
The integration thus reduces to finding gauge configurations sampled according to
the weight from the gauge action and the fermion determinant(s). On each such
gauge configuration one computes the quark propagators from some source point
to all other points of the lattice by inverting the Dirac operator matrix
$(D+m)$. Combination of the quark propagators gives the contribution to
the correlation functions. This is repeated for as many gauge configurations as
possible.

Projecting to $\vec p =0$ simplifies the analysis, since the energy of the
hadron at rest is just its mass. Due to the usually chosen boundary conditions
a meson propagator is periodic and will have contributions from propagation
forward as well as backward in Euclidean time,
\be
C_\pi(\vec p=0,t)\sim \E^{-a m_\pi t}+\E^{-a m_\pi (T-t)}
\;,
\ee
where $t$ denotes the time in lattice units and $T$ is the extent of the lattice
in temporal direction. The above behavior is only asymptotically, for large
temporal distances. For smaller values of $t$ higher excitations will modify the
functional form. This is one of the issues to be discussed later.  Computing the
pion mass leads to results like the GMOR relation $m_\pi^2 \propto m_q$, which
has been verified in many lattice  calculations (see, e.g., \cite{BGR04}).

The construction of the interpolating field operators is inspired by the heavy
quark limit.  Remember, that site fields correspond to fields implicitly averaged
over lattice distances of $a$.  Almost any operator with the correct quantum
numbers should do, although bad choices may not couple strongly to the physical
state. The correlation function will have a larger amplitude  if the overlap with
the physical state is better and thus the signal improves.

There is some freedom for choosing the hadron interpolating fields. Simple examples like
the nucleon-type operator
\be
N^{(i)}=\epsilon_{abc} \,\Gamma^{(i)}_1 \,u_a\,
\left( u^T_b \,\Gamma^{(i)}_2\, d_c - d^T_b \,\Gamma^{(i)}_2\, u_c\right)\FD
\ee
with the choices
\begin{center}
\begin{tabular}{r|cc}
& $\Gamma^{(i)}_1$ & $\Gamma^{(i)}_2$ \\
\hline
$i=1$& $1$ & $C \gamma_5$ \\
$i=2$& $\gamma_5$ & $C$ \\
$i=3$& $\I$ & $C \gamma_4\gamma_5$ \\
\end{tabular}
\end{center}
are important representatives ($C$ denotes the charge conjugation operator in
Dirac space). For such operators one usually fixes the source at an arbitrarily
chosen lattice site (like the origin) and has to compute the quark propagator 
$(D+m)^{-1}_{0x}$ from this point to all other points on the
lattice. This allows to compute the hadron propagator from this source to any
other sink position.  Averaging over the sink operators in some timeslices then projects
to vanishing sink 3-momentum. Summing over many gauge configurations amounts to
averaging out all but the $\vec p=0$ component of the correlator.

Next to such ``local'' operators, where all quarks live on the same lattice
site, one can use extended operators combining quarks at different, nearby
points  (cf., \cite{EdFlJo07} and references therein). This way one can utilize
the lattice symmetries to construct representations of the cubic group. Such
combinations correspond to certain combinations of angular momentum
representation in the continuum. Such an approach involves the computation of
more quark propagators, one for each sink position. 

Another way to improve the signal is to use some type of smearing. One can
locally average the gauge links (gauge configuration smearing). This should not
affect the correlation function over larger time distance. One can, however,
instead of point-like quark sources also use modified ones that are smeared
over some region. In the BGR collaboration we used Gaussian smearing which
refers to the Gaussian shape of the source
\cite{BuGaGl04aBuGaGl05c,BuGaGl06,BuGaGl06b}. Here one needs to compute quark
propagators for each type of sink, e.g, for a narrow and a widely smeared sink.
One then combines differently smeared quark propagators to hadron operators.
This approach improves the statistical quality of the correlation functions
significantly.

\subsection{\it The problem with excited states}

The real problem with excited states is that they are unstable, they decay. They are
no asymptotic  states of the theory and thus we cannot go to large time
distances to identify them, we have to rely on other methods. Fortunately the
finiteness of the lattice comes to the rescue. On a finite lattice the energy
spectrum is not continuous but discrete. By studying the volume dependence of
the spectrum one can learn about  the scattering phase shift and other
properties of the scattering amplitude. This has been studied in detail
\cite{Mi85,Lu86a,LuWo90,Lu91} and first attempts to use this approach in
full QCD simulations have been published \cite{McMi03,AoFuIs}.
The central problem is to reliably determine the energy spectrum in lattice
calculations.

In quenched simulations life is somewhat simpler:
The states cannot decay since there is no creation of
quark--anti-quark loops from the vacuum. The spectrum in some channel
with definite  quantum numbers is directly related to masses of excited states.
There is, however, a shadow of dynamical quarks due to the quark-lines turning
around and running backwards for some time (cf., the 2nd graph in
Fig.\ \ref{fig:ghost}). This may lead to effects mimicking
intermediate 2-particle states. We come back to that problem, which leads to
quenched ghosts, later on.

A correlation function between hadronic operators involves a sum over  a tower
of intermediate states, as can be seen inserting a complete set of (physical)
mass eigenstates. Let us assume $X(t)$ denotes the operator with vanishing
3-momentum at Euclidean time $t$, then the correlator is
\bea\label{eq:qmintermedstates}
\langle X(t) \overline X(0)\rangle
&=&\sum_n\; \langle X(t)|n\rangle \,\E^{-m_n \,t}\,\langle n|\overline X(0)\rangle
\\
&\sim&
a_1 \,\E^{-m_1 \,t}
+a_2 \,\E^{-m_2 \,t}
+a_3 \,\E^{-m_3 \,t}+\ldots
\eea
The leading term with the smallest mass is the ground state, which will be
dominantly seen at asymptotically large $t$. The other states are the
excitations which we are interested in. They will be observed at smaller $t$
and  dominate the ground state signal. In the Monte Carlo calculation we have to
somehow disentangle these states. Simply fitting to a sum of exponentials is
usually unstable and needs  extremely good statistics. Several better methods
have been used.

The {\em Bayesian analysis} is based on ideas from Ref.\ \cite{LeClDa01} and has
been used in extracting excited hadrons in a sequence of papers 
\cite{LeDoDr03ChDoDr04,DoDrHo05}. This is a stepwise procedure, where on
starts by extracting the ground state mass from larger $t$ and then uses that
value as a prior for getting estimates of the higher lying states at smaller
$t$.

Another method combines correlation function values from different  time slices
such as to ``subtract'' the leading exponential behavior \cite{GuPaSi04}; the
combination leads to rapidly decaying correlators.

Another approach used was to reconstruct the spectral density (as a continuous
function) with the {\em maximum entropy method} \cite{SaSaHa05}. This method
results in a probability function for the masses. One seems to need high
(statistical) quality data for a successful application.

In the BGR collaboration \cite{BuGaGl04aBuGaGl05c,BuGaGl06,BuGaGl06b} and also
in the LHPC collaboration and the Adelaide group
\cite{MeBiBo03ZaLeWi03,LaHeKa07,EdFlJo07, BaEdFl} the so-called {\em variational method} is used. 
The idea is straightforward. In functional analysis  a function can be expanded
in a series of independent basis functions.  In the same way the unknown
operator of the physical state can be built from a sum over basis
operators. Contrary to simple continuum mathematics, we do not know how to
orthogonalize our interpolating field operators.  We can, however, study
correlation matrix elements between the set of interpolating operators,
\be
C_{ij}(t)=\langle X_i(t)\overline X_j(0)\rangle\FD
\ee
Like in the spectral representation \eq{eq:qmintermedstates} these correlators
will involve implicitly sums over the physical eigenstates. In
\cite{Mi85,LuWo90} it was shown, that the composition of the physical operators
in terms of the interpolating operators can be recovered by solving the
generalized eigenvalue problem,
\be
C(t)\,u^{(n)} = \lambda^{(n)}(t) C(t_0) \,u^{(n)} \FD
\ee
The eigenvalues will, in leading order, behave like single exponentials and
allow us to recover the energy (or for vanishing 3-momentum, the mass) values of
the intermediate  physical states:
\be
\lambda^{(n)}(t)\approx \E^{-E_n\,(t-t_0)}
\left(1+\mathcal{O}\left(\E^{-\Delta E_n\,(t-t_0)}\right)\right)\FD
\ee
Here  $\Delta E_n$ denotes the energy difference to the next higher state. The
eigenvectors give an idea on the composition of the physical states and provide
a fingerprint of the state.

There are some caveats. The number of possible interpolating fields is limited
due to resources. Also, including too many of them adds statistical
noise to the correlation matrix affecting the reliability of the
diagonalization.  Therefore it pays off to find ``good'' interpolators. One
has to stress, however, that there is no a priori bias for the composition of
the physical states, except for the choice of the basis set of
interpolating fields. 

Note that the diagonalization is done for each time distance $t$ independently.
Thus it is important to study the consistent identification of each state by 
following the eigenvectors. In practice one tries to find stability plateaus,
such that the energy values and the eigenvectors
remain constant over a range of $t$-values.

\begin{figure}[tb]
\begin{center}
\epsfig{file=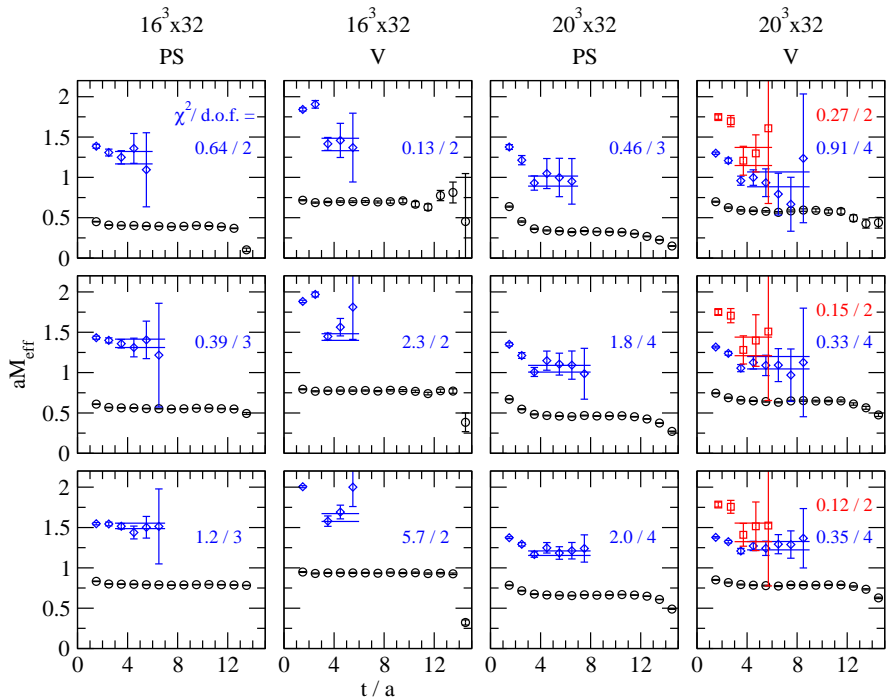,scale=0.5,angle=270,clip=}
\caption{Effective-mass plots for pseudoscalar (PS) and vector (V) ground state
mesons and first excitations for different size lattices (Fig. from
\cite{BuGaGl06}). \label{fig:eigenvectors}}
\end{center}
\end{figure}

One can define effective masses
\be
M_\srm{eff}\left(t+\ot\right)=\ln\left(\lambda(t)/\lambda(t+1)\right)
\ee
for each eigenvalue $\lambda_n(t)$ and plot these like in Fig.\
\ref{fig:eigenvectors} in order to get an idea on the plateau range. The fit to
the final mass value is then done with an exponential fit over all points in the
plateau range. The number of eigenvalues to
be trusted is of course smaller than the order of the correlation matrix. It
can be determined by studying the effect when enhancing or reducing the set of
interpolators. The typical experience is that the lower  2/3 of the energy
levels are reliable.

Combining different Dirac structures and differently smeared quark sources to
hadron interpolators gives a large  basis set \cite{BuGaGl06b,BuGaGl06}. Not all
of them bring in new content, though, and one tries to minimize their number in
order to improve the statistical quality of the diagonalization result.
Physical intuition guides the construction; a recent inclusion of derivative
sources led to some improvement \cite{GaGlLa07}.

\subsection{\it Further issues}

Of course it would be interesting and important to study, e,g., the scalar
isoscalar $0^{++}$ meson channel ($f_0$, also called $\sigma$ or $\epsilon$
depending on your favorite edition of the particle data booklet). This is
rarely done so far. The reason is that the propagator for such a $\ubar\, u$
state involved disconnected, backtracking loops. In order to project to zero
momentum one would need such loops for all $L^3$ sites of a timeslice, and thus
compute $L^3$ quark propagators. Otherwise the signal would be extremely noisy.
This is very costly and therefore one tries to invent methods (like low-mode
averaging \cite{DeSc04a}) to  improve the situation. Most groups avoid this
problem by considering only hadron correlation functions which do not involve
disconnected terms.

In principle 2-particle channels may obscure the single particle signal. Even
in quenched calculations there may be quasi-2-particle states due to $Z$-shaped
quark lines. A possible identification method is to study the volume dependence
of the spectral weight \cite{MaAlCh06}. For 2-particle states in the rest frame
there may be relative spatial momenta, which then have a volume dependence  via
the spectral relation $E=2\sqrt{m^2+p^2}$, where $p$  may be multiples of the
minimum lattice momentum $2\pi/L$. Also studying the operators with
non-vanishing total momentum allows one to distinguish single particle states from
2-particle states.

Finite volume quenches the states, leading to higher mass values for hadrons with
large physical size. This and another effect related to light states (the pion)
running around the usually periodic spatial lattice has to be considered as well.

As has been pointed out \cite{BaDuEi01}, in the quenched situation the $Z$-graph
like contributions may  mimic ghost states like an $\eta'$, leading to negative
contributions to the propagator, in particular towards smaller quark masses.
This has been verified in both, baryonic and mesonic channels
\cite{DoDrHo05,MaAlCh06,BuGaGl06a}. The functional behavior of this
contribution deviates from a simple exponential. The term has to be either
removed with an ansatz for its shape \cite{DoDrHo05,MaAlCh06} or by isolating
it in the variational method \cite{BuGaGl06a}.

The approach towards smaller quark masses is cumbersome. In the quenched case
there are additional singularities (quenched chiral logs). In the case with
dynamical quark decay channel are opening and the identification of states like
the $\rho$ becomes more complicated.

\section{Some results}

\subsection{\it Quenched results}

In a quenched simulation \cite{BuGaGl06,BuGaGl06b} we have generated gauge
configurations with the L\"uscher-Weisz gauge action  \cite{LuWe85} on
lattices of size $16^3\times 32$ and $20^3\times 32$. The gauge coupling was
chosen to work at lattice spacings between 0.12 and 0.15 fm, corresponding to a
spatial lattice size around 2.4 fm. We used two mass-degenerate light quarks
(u, d) and one heavier s-quark with its mass parameter adjusted to give the
correct K-meson mass.

One of the open puzzles is the behavior of the first excitation in the positive
parity  nucleon channel, the so-called Roper resonance N(1440). In quantum
mechanics and in the heavy quark limit the positive and negative parity
excitations come with alternate parities $(+\,-\,+\,-)$. The  experimentally
observed states show a different order: N(940), N(1440), N(1535) and N(1650)
corresponding to $(+\,+\,-\,-)$. Since in the lattice simulations we proceed from
high to low quark masses we expect a level crossing at some point. So far, in the
regions accessible to studies, only one group has seen such a crossing at smaller
quark masses \cite{DoDrHo05}, albeit with large error bars. In that study 
smaller pion masses were accessible due to using the overlap action; the analysis
was of the discussed Bayesian type. Other groups, including ours, identify the
Roper state at high masses but its behavior is inconspicuous towards lower quark
masses and no drastic change of slope is observed. Clearly there are two
concerns: (a) the finite volume may quench the excited states enhancing their
energy or (b) dynamical fermions may be important when approaching the chiral
limit. No such deviation is noticed for the negative parity sector.

\begin{figure}[tb]
\begin{center}
\epsfig{file=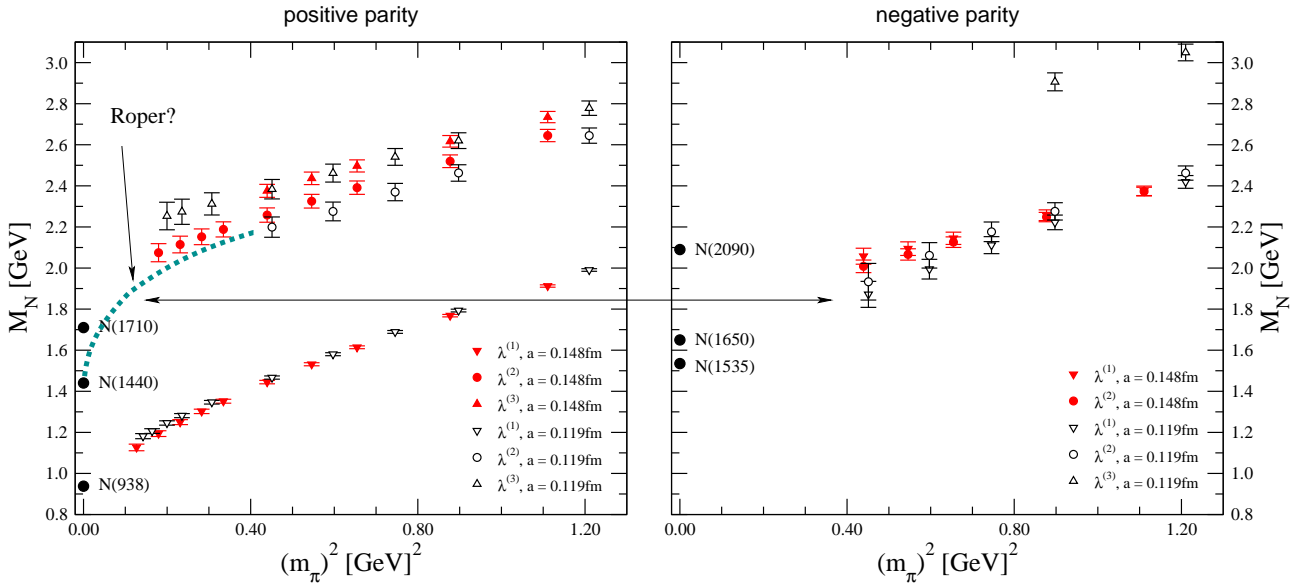,scale=0.5,clip}
\caption{Quenched results for the positive and negative parity  nucleons from
Ref.\ \cite{BuGaGl06b}. The Roper states N(1440) and the N(1710) come out too high and there is no
indication of level crossing. \label{fig:nucleonroper}}
\end{center}
\end{figure}

Chiral perturbation theory \cite{We79GaLe84} (cf., the talks
\cite{ThomasErice,WeiseErice}) is a systematic expansion around the chiral
limit. The effective Lagrangian introduces parameters, which can be determined
by comparing with experiment (or lattice results) \cite{Me05,Bi07}. Expansion
of the nucleon mass involves terms $\mathcal{O}(m)$, $\mathcal{O}(m^2)\propto
m_\pi$, and so on, (the exact form depending on whether one uses quenched ChPT
\cite{LaSh96} or not). Excited states bring in new scales and new assumptions.
In such an expansion around $M_\pi=0$  the Roper and the ground state nucleon
behave similarly \cite{BoBrMe06}.

Fig. \ref{fig:quebarspec} exhibits the results of a simple extrapolation to the
physical point \cite{BuGaGl06b} and we have discussed some conclusions in the
introduction. The agreement of the negative parity extrapolations is remarkable
and led us to suggest to look for a negative parity state $\Omega(1970)$ and
two $\Xi$ states near 1780 MeV.

\begin{figure}[tb]
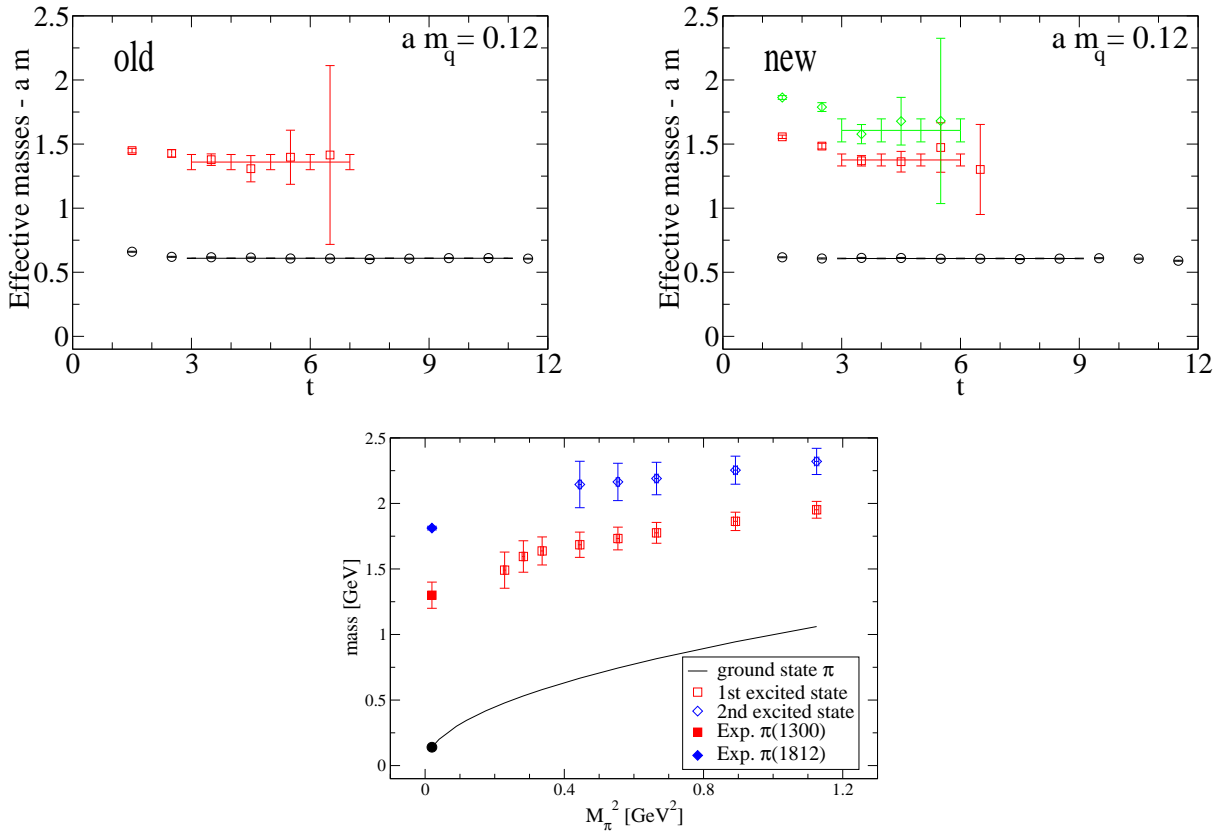

\begin{center}
\epsfig{file=figs/masses_100111000000_m0.12.eps,scale=0.35,clip}\hspace{32pt}
\epsfig{file=figs/masses_100001001100_m0.12.eps,scale=0.35,clip}\vspace{12pt}\\
\epsfig{file=figs/massplot_100101001001.eps,scale=0.3,clip}
\caption{The upper two plots (from \cite{GaGlLa07}) compare the results of an
analysis of the pseudoscalar pion channel without (old) and with (new)
derivative quark sources. The lower plot shows the obtained values for several
valence quark masses. The ground state pion curve is plotted just to guide the
eye.\label{fig:derivpions}}
\end{center}
\end{figure}

In the meson sector the discussed variational technique allowed us to clearly
identify the first pseudoscalar and vector  excitations. A
recent extension of that study includes covariant (nearest neighbor) derivative
quark sources, which introduce another set of interpolators. Indeed first
results show that this allows one to obtained higher excitations in certain
channels \cite{GaGlLa07}. Fig.\ \ref{fig:derivpions} demonstrates this
improvement for the pseudoscalar channel.

\subsection{\it Dynamical fermions}

To include dynamical fermions in the Monte Carlo simulation means to include
the determinant of the  Dirac operator matrix as a weight factor in the
generation of the gauge configurations. This is done by utilizing the relation
between Grassmann and bosonic integration:
\be
\mathrm{det} ( D\, D^\dagger)
= \int [\mathcal{D} \psi \mathcal{D} \psibar] 
\exp(-\psibar( D\, D^\dagger) \psi)
= \int [\mathcal{D} \phi \mathcal{D} \phi^\dagger] 
\exp(-\phi^\dagger ( D\, D^\dagger)^{-1} \phi)\FD
\ee
The bosonic fields $\phi$ and $\phi^\dagger$ have the same degrees of freedom as
the fermions and are called pseudo\-fermions. In a simulation the change of the
fermion determinant due to suggested changes of the gauge fields thus can be
expressed through the bosonic variables. Due to the inversion of the matrix the
effective bosonic action is very non-local and the simulation becomes
expensive. The standard method is to evolve the gauge field in a molecular
dynamics algorithm for some time and then correct for accumulated errors by a
Monte-Carlo accept-reject step. This is called one trajectory in this so-called
Hybrid Monte-Carlo (HMC) algorithm \cite{DuKePe87}.

In the molecular dynamics part one has to compute the inverse and the 
derivative of the Dirac operators with regard to the changed gauge fields quite
often. This is time consuming, in particular for more complicated actions with
many terms like the CI action.

\begin{figure}[tbp]
\begin{center}
\epsfig{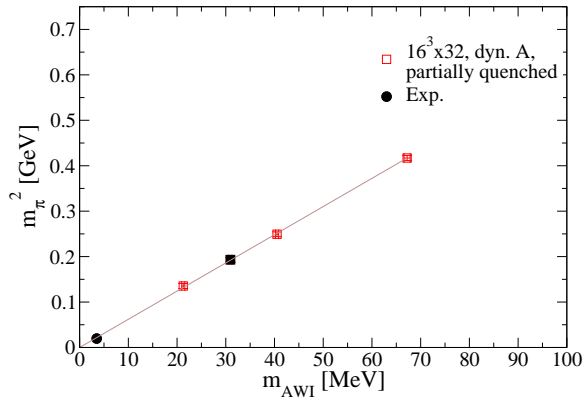}
\end{center}
\caption{\label{fig:gmor} Pion mass squared vs. the quark mass for run A. The
data point marked by the filled square (2nd from below) is the value where the valence quark and
the sea quark masses agree, the other three points are partially quenched,
i.e., the valence quark mass differs from the sea quark mass. In that plot the
quark mass has been determined with help of the axial Ward identity (AWI) from
the ratio of a correlators involving the derivative of the axial current and
the pseudoscalar field, thus called $m_\srm{AWI}$. }
\end{figure}

In \cite{LaMaOr05c} we have discussed our implementation of the HMC  algorithm
for the CI action and shown first results on $12^3\times 16$ lattices.
Meanwhile we have been extending the study to larger $16^3\times 32$ lattices
with a lattice spacing around $0.154$ fm \cite{FrGaLa07}. We use the
L\"uscher-Weisz gauge action, stout smearing and two mass-degenerate light
quarks. The molecular dynamics trajectory has 100 steps for one unit of
HMC-time and we analyze every 5th configuration to reduce autocorrelation. Up
to the moment of the Erice meeting we have produce $\mathcal{O}(50)$
independent configurations for two combinations of mass and lattice spacing
(cf., Table \ref{tab:dyn}). 

\begin{table}[h]
\begin{center}
\begin{tabular}{rrrrrrrr}
run & $\beta_\srm{LW}$ & $c_0$  & $n_\srm{conf}$ & $n_\srm{meas}$ 
& $a$ [fm]&  $M_\pi$ [MeV]  \cr
\hline 
A & 4.65 & -0.06 & 425 & 65 & 0.153(1) & 463(4)\cr
B & 4.70 & -0.05 & 350 & 50 & 0.154(1) & 507(5) \cr
\hline
\end{tabular}
\end{center}
\caption{Parameters of the two runs discussed: run sequence, gauge coupling, 
bare mass parameter $c_0$, number of configurations $n_\srm{conf}$, number of
configurations analyzed $n_\srm{meas}$, lattice spacing $a$  assuming that the
Sommer parameter is $r_0=0.48$ fm, pion mass.}
\label{tab:dyn}
\end{table}

\begin{figure}[bp]
\begin{center}
\epsfig{file=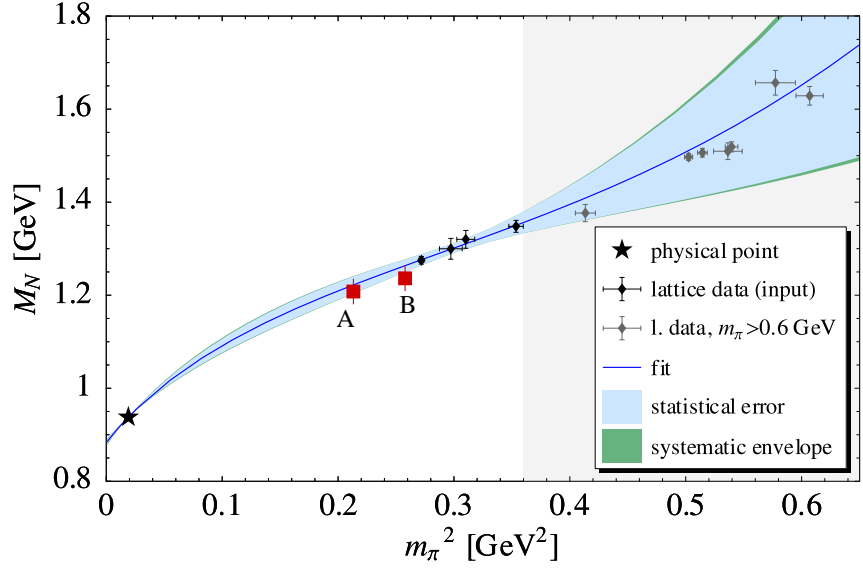,scale=0.85,clip}
\end{center}
\caption{\label{fig:nucleon_chpt} Our results for the CI-action HMC simulation
(full squares) compared to the ChPT analysis error band and lattice data as
discussed in Ref.\ \cite{PrMuWo06}  (original figure from that reference, data
as quoted \cite{AlBaGo04,OrLiSc05}). }
\end{figure}

\begin{figure}[tbp]
\begin{center}
\epsfig{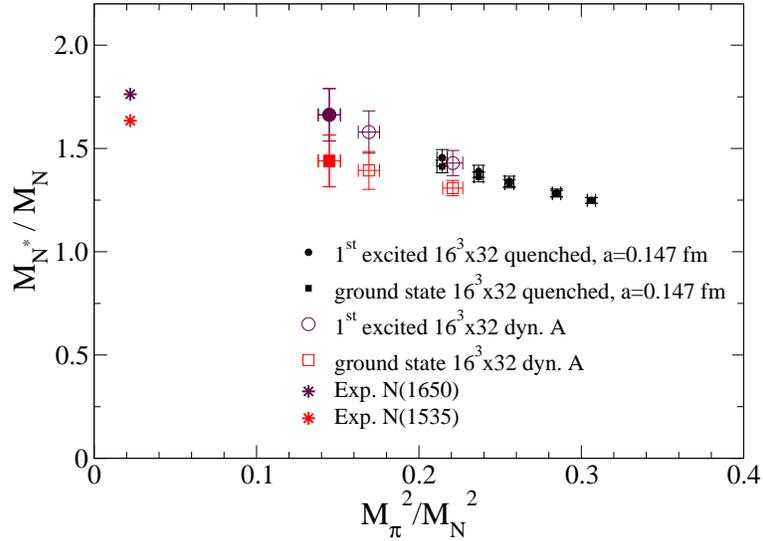}
\end{center}
\caption{We can identify the two lowest lying negative parity nucleons
\cite{FrGaLa07} which are compatible with a smooth extrapolation to the
experimental values. \label{fig:dynnegparnucleons}
}
\end{figure}

In a simulation with dynamical quarks the sea quark mass agrees with the valence
quark mass; thus one has just one data point, unlike a quenched simulation
where one shows many different valence quark masses. In an attempt to better
understand the difference one  relies also in dynamical quark simulations on
so-called partially quenched data points, where the valence quark mass differs
from the sea quark mass.  Fig.\ \ref{fig:gmor} shows such a result, where
indeed the partially quenched values for the pion mass together with the only
``correct'' dynamical point (second square from the left) extrapolate nicely to
the chiral limit. This is another example of the good agreement of lattice
results with the (leading order ChPT) Gell-Mann--Oakes--Renner relation
\cite{GeOaRe68},
\be \label{eq:GMOR}
f_\pi^2\, M_\pi^2=- 2\, m \,\langle \overline q q\rangle \FD
\ee

Comparing our results for the nucleon mass with, e.g., the chiral analysis of
Ref.\ \cite{PrMuWo06}, we find (see Fig\ \ref{fig:nucleon_chpt}) excellent
agreement with the chiral interpolation between the experimental nucleon mass
and other lattice results.

Our results for dynamical quarks brought no surprises for the negative parity
nucleons (Fig. \ref{fig:dynnegparnucleons}). We continue to find two nearby
states extrapolating to the physical value towards the smaller quark masses,
behavior and values very similar to the quenched situation.

\begin{figure}[tb]
\begin{center}
\epsfig{file=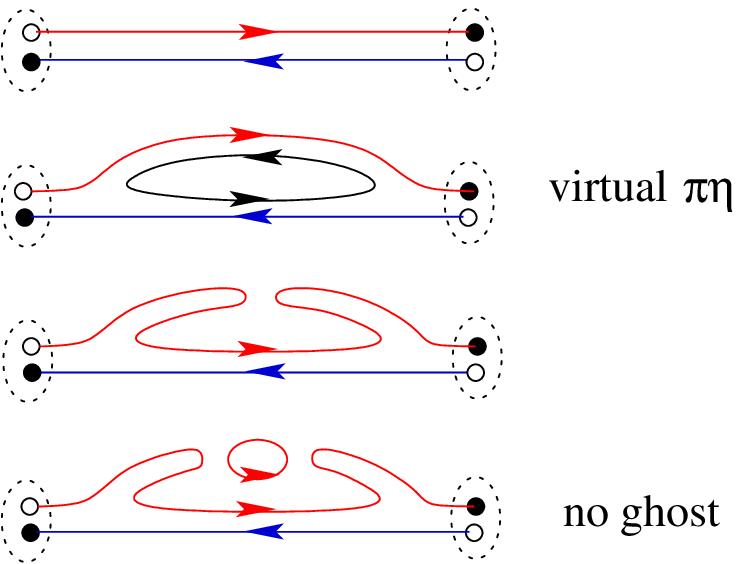,scale=0.8,clip}\hspace{2cm}
\hbox{\epsfig{file=figs/ghost_plot_a0.eps,scale=0.7,clip}}
\end{center}
\caption{Left: Graphs contributing to the isovector channel; the 2nd and the
4th are missing in the quenched approximation. Right: One of the
three leading eigenvalues in the quenched $a_0$-correlation function shows
the ghost contribution \cite{BuGaGl06a}.
\label{fig:ghost} }
\end{figure}

A particularly problematic mesonic channel is that of the isovector, scalar meson
$a_0$. In quenched calculations this channel is plagued by an artifact. In
Fig.\ \ref{fig:ghost} the quark lines contributing to the propagator are
indicated. In the quenched case only the 1st and 3rd diagrams contribute. As
discussed earlier, the 3rd diagram gives,  for small quark masses, rise to a
virtual $\pi\eta'$ ghost state, which adds a negative contribution to the total
propagator \cite{BaDuEi01,DoDrHo05,MaAlCh06}. The right-hand side plot (from Ref.\
\cite{BuGaGl06a}) shows such a contribution; it decouples in the variational
approach, but due to limited statistics this decoupling is not perfect at
smaller masses, rendering the extrapolation  of the $a_0$-mass towards smaller
quark masses problematic.

\begin{figure}[tb]
\begin{center}
\epsfig{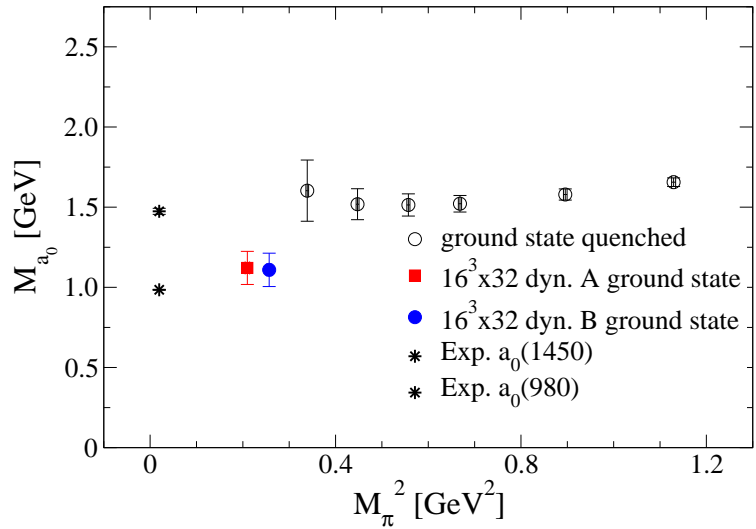}
\end{center}
\caption{Whereas in the quenched case we (like other authors) obtain a $a_0$-mass
which points towards the $a_0$(1450), the dynamical results seem to indicate a
low lying state compatible with $a_0$(980). \label{fig:a0dyn}
}
\end{figure}

In quenched calculations the $a_0$-mass was always seen too high, compatible
with the first excitation $a_0$(1450). The behavior in the dynamical runs came
as a surprise. We now see the lowest mass compatible with the ground state
$a_0$(980) (see Fig.\ \ref{fig:a0dyn}). 

In full QCD the 4th diagram in Fig.\ \ref{fig:ghost} removes the  ghost
behavior, the $\pi\eta$ channel now couples as a physical state. Also, since we
have relatively high quark masses $\mathcal{O}(30\ldots 40)$ MeV, the now
possible quark vacuum loops  cannot really distinguish between light and heavy
(strange) quark contributions. Since this 2-particle channel mixes with the
$a_0$ state we will have to study its momentum, volume and quark mass
dependence to make sure that the observed energy value corresponds really to a
single particle mass \cite{McMi06aMc07a}.

\section{Summary and conclusion}

Although lattice calculations have gone a long way, there are still hard
challenges ahead. Excited studies for full QCD requires systematic exploration
of volume dependence, quark mass dependence and scaling towards the continuum
limit. First steps in determining properties of excited states (like decay
constants) are under progress \cite{McMi06Mc07}.  The puzzle of the Roper
state and its level crossing should be clarified. The future is wide open.

\section{Acknowledgment}

I want to thank the organizers for the enjoyable meeting. Special thanks go to
all my collaborators in the BGR collaboration for many years of  challenging and
fruitful cooperation, in particular to T. Burch, R. Frigori, C. Gattringer, 
L.Y. Glozman, C. Hagen, P. Hasenfratz, D. Hierl, Ph. Huber,  M. Limmer, P.
Majumdar, T. Maurer, D. Mohler, F. Niedermayer, W. Ortner, and A. Sch\"afer.


\begin{thebibliography}{10}
\itemsep -2pt 
\bibitem{BuGaGl06b}
T. Burch et~al.,
{\em  Phys. Rev.} D 74 (2006) 014504, hep-lat/0604019.

\bibitem{BuGaGl06}
T. Burch et~al.,
{\em  Phys. Rev.} D 73 (2006) 094505, hep-lat/0601026.

\bibitem{Gl07}
L.Y. Glozman,
{\em  Phys. Rept.} 444 (2007) 1, hep-ph/0701081.
%%CITATION = HEP-PH/0701081;%%

\bibitem{Wi74}
K.G. Wilson,
{\em Phys. Rev.} D 10 (1974) 2445.

\bibitem{Cr80}
M. Creutz,
{\em Phys. Rev.} D 21 (1980) 2308.
%%CITATION = PHRVA,D21,2308;%%

\bibitem{We79GaLe84}
S. Weinberg,
{\em Physica} A 96 (1979) 327.
J. Gasser and H. Leutwyler,
{\em Ann. Phys.} 158 (1984) 142.

\bibitem{NiNi}
H.B. Nielsen and M. Ninomiya,
{\em Phys. Lett.} B 105 (1981) 219; {\em Nucl. Phys.} B 185 (1981) 20;
{\em Nucl. Phys.} B 193 (1981) 173.

\bibitem{GiWi82}
P.H. Ginsparg and K.G. Wilson,
{\em Phys. Rev.} D 25 (1982) 2649.

\bibitem{HaNi94HaHaNi05}
P. Hasenfratz and F. Niedermayer,
{\em Nucl. Phys.} B 414 (1994) 785.
A. Hasenfratz, P. Hasenfratz and F. Niedermayer,
{\em Phys. Rev.} D 72 (2005) 114508, hep-lat/0506024.
%%CITATION = HEP-LAT 0506024;%%

\bibitem{Ne}
H. Neuberger,
{\em Phys. Lett.} B 417 (1998) 141; ibid. B 427 (1998) 353.

\bibitem{Lu98}
M. L{\"u}scher,
{\em Phys. Lett.} B 428 (1998) 342.

\bibitem{Ka92}
D.B. Kaplan,
{\em Phys. Lett.} B 288 (1992) 342, hep-lat/9206013.

\bibitem{FuSh95}
V. Furman and Y. Shamir,
{\em Nucl. Phys.} B439 (1995) 54, hep-lat/9405004.
%%CITATION = HEP-LAT 9405004;%%

\bibitem{Ga01aGaHiLa00}
C. Gattringer,
{\em Phys. Rev.} D 63 (2001) 114501, hep-lat/0003005.
C. Gattringer, I. Hip and C.B. Lang,
{\em Nucl. Phys.} B597 (2001) 451, hep-lat/0007042.
%%CITATION = HEP-LAT 0007042;%%

\bibitem{BGR04}
C. Gattringer et~al. (BGR (Bern-Graz-Regensburg) collaboration),
{\em Nucl. Phys.} B 677 (2004) 3, hep-lat/0307013.

\bibitem{EdFlJo07}
R.G. Edwards et~al. (LHPC collaboration),
{\em PoS} LATTICE2007 (2007) 108, arXiv:0710.3571 [hep-lat].
%%CITATION = ARXIV:0710.3571;%%

\bibitem{BuGaGl04aBuGaGl05c}
T. Burch et~al.,
{\em Phys. Rev.} D 70 (2004) 054502, hep-lat/0405006.
%%CITATION = HEP-LAT 0405006;%%
T. Burch et~al.,
{\em PoS} LAT2005 (2005) 097, hep-lat/0509086.

\bibitem{Mi85}
C. Michael,
{\em Nucl. Phys.} B 259 (1985) 58.

\bibitem{Lu86a}
M. L{\"u}scher,
{\em Commun. Math. Phys.} 105 (1986) 153.

\bibitem{LuWo90}
M. L{\"u}scher and U. Wolff,
{\em Nucl. Pbys.} B 339 (1990) 222.

\bibitem{Lu91}
M. L{\"u}scher,
{\em Nucl. Phys.} B 354 (1991) 531;
ibid. B 364 (1991) 237.

\bibitem{McMi03}
C. McNeile and C. Michael,
{\em Phys. Lett.} B 556 (2006) 177, hep-lat/0212020.

\bibitem{AoFuIs}
S. Aoki et~al.,
{\em PoS} LAT2006 (2006) 110, hep-lat/0610020.
S. Aoki et~al.,
{\em Phys. Rev. } D 76 (2007) 094506, arXiv:0708.3705v1.


\bibitem{LeClDa01}
G.P. Lepage et~al.,
{\em Nucl. Phys. Proc. Suppl.} 106 (2002) 12, hep-lat/0110175.
%%CITATION = HEP-LAT/0110175;%%

\bibitem{LeDoDr03ChDoDr04}
F. Lee et~al.,
{\em Nucl. Phys.} Proc. Suppl. 119 (2003) 296, hep-lat/0208070.
%%CITATION = HEP-LAT/0208070;%%
Y. Chen et~al.,
\newblock {T}he {S}equential {E}mpirical {B}ayes {M}ethod: {A}n {A}daptive
  {C}onstrained-{C}urve {F}itting {A}lgorithm for {L}attice {QCD},
hep-lat/0405001.

\bibitem{DoDrHo05}
S.J. Dong et~al.,
{\em Phys. Lett.} B 605 (2005) 137, hep-ph/0306199.


\bibitem{GuPaSi04}
D.~Guadagnoli, M.~Papinutto and S.~Simula,
{\em Phys.\ Lett.} B 604 (2004) 74, hep-lat/0409011 .
%%CITATION = HEP-LAT 0409011;%%

\bibitem{SaSaHa05}
K. Sasaki, S. Sasaki and T. Hatsuda,
{\em Phys. Lett.} B 623 (2005) 208, hep-lat/0504020.
%%CITATION = HEP-LAT/0504020;%%

\bibitem{MeBiBo03ZaLeWi03}
W. Melnitchouk et~al.,
{\em Phys. Rev.} D 67 (2003) 114506, arXiv:hep-lat/0202022v3.
J. M. Zanotti et~al.,
{\em Phys. Rev.} D68 (2003) 054506, hep-lat/0304001.
%%CITATION = HEP-LAT/0304001;%%

\bibitem{LaHeKa07}
B.G. Lasscock et~al.,
{\em Phys. Rev.} D 76 (2007) 054510, arXiv:0705.0861

\bibitem{BaEdFl}
S. Basak et~al.,
{\em Phys. Rev.} D 72 (2005) 094506, hep-lat/0506029;
ibid. {\em Phys. Rev.} D 72 (2005) 074501hep-lat/0508018, .
{\em Nucl. Phys.} (Proc. Suppl.) 153 (2006) 242, hep-lat/0601034.
{\em PoS} LAT2006 (2006) 197, hep-lat/0609072.
{\em Phys. Rev.} D 76 (2007) 074504, arXiv:0709.0008.

\bibitem{GaGlLa07}
C. Gattringer et~al.,
{\em PoS} LATTICE 2007 (2007) 123, arXiv:0709.4456.
%%CITATION = ARXIV:0709.4456;%%

\bibitem{DeSc04a}
T. DeGrand and S. Schaefer,
{\em Nucl. Phys. Proc. Suppl.} 140 (2005) 296, hep-lat/0409056;
{\em Comput. Phys. Commun.} 159 (2004) 185, hep-lat/0401011
%%CITATION = HEP-LAT/0409056;%%

\bibitem{MaAlCh06}
N. Mathur et~al.,
\newblock {S}calar {M}esons $a_0(1450)$ and $\sigma(600)$ from {L}attice {QCD},
\newblock hep-ph/0607110.

\bibitem{BaDuEi01}
W.A. Bardeen et~al.,
{\em Phys. Rev.} D65 (2001) 014509, hep-lat/0106008.
%%CITATION = HEP-LAT/0106008;%%

\bibitem{BuGaGl06a}
T. Burch et~al.,
{\em  Phys. Rev. } D 73 (2006) 017502, hep-lat/0511054.
%%CITATION = HEP-LAT 0511054;%%

\bibitem{LuWe85}
M. L{\"u}scher and P. Weisz,
{\em Commun. Math. Phys.} 97 (1985) 59;
ibid. 98 (1985) 433.

\bibitem{ThomasErice}
A. Thomas,
\newblock {H}adron structure and quark models - the ``spin crisis'',
\newblock Contribution to these proceedings, 2007.

\bibitem{WeiseErice}
W. Weise,
\newblock {C}hiral extrapolations of nucleon properties,
\newblock Contribution to these proceedings, 2007.

\bibitem{Me05}
U.G. Mei{ss}ner,
{\em PoS} LAT2005 (2006) 009, hep-lat/0509029.
%%CITATION = HEP-LAT/0509029;%%

\bibitem{Bi07}
J. Bijnens,
{\em PoS} LATTICE2007 (2007) 004, arXiv:0708.1377 [hep-lat]
%%CITATION = ARXIV:0708.1377;%%

\bibitem{LaSh96}
J.N. Labrenz and S.R. Sharpe,
{\em Phys. Rev.} D 54 (1996) 4595, hep-lat/9605034.

\bibitem{BoBrMe06}
B. Borasoy et~al.,
{\em Phys. Lett.} B 641 (2006) 294, hep-lat/0608001.
%%CITATION = HEP-LAT/0608001;%%

\bibitem{DuKePe87}
S. Duane et~al.,
{\em Phys. Lett.} B 195 (1987) 216.
%%CITATION = PHLTA,B195,216;%%

\bibitem{LaMaOr05c}
C.B. Lang, P. Majumdar and W. Ortner,
{\em Phys. Rev.} D 73 (2006) 034507, hep-lat/0512014.
%%CITATION = HEP-LAT 0512014;%%

\bibitem{FrGaLa07}
R. Frigori et~al.,
{\em PoS} LATTICE2007 (2007) 114, arXiv:0709.4582v1 [hep-lat].

\bibitem{PrMuWo06}
M. Procura et~al.,
{\em Phys. Rev.} D 73 (2006) 114510, hep-lat/0603001.
%%CITATION = HEP-LAT/0603001;%%

\bibitem{AlBaGo04}
A. Ali Khan et~al.,
{\em Nucl. Phys.} B 689 (2004) 175, hep-lat/0312030. 

\bibitem{OrLiSc05}
B. Orth, T. Lippert and K. Schilling,
{\em Phys. Rev.} D 72 (2005) 014503, hep-lat/0503016.

\bibitem{GeOaRe68}
M. Gell-Mann, R.J. Oakes and B. Renner,
{\em Phys. Rev.} 175 (1968) 2195.

\bibitem{McMi06aMc07a}
C. McNeile and C. Michael,
{\em Phys. Rev.} D 74 (2006) 014508, arXiv:hep-lat/0604009.
C. McNeile,
{\em PoS} LATTICE2007 (2007) 019, arXiv:0710.0985.

\bibitem{McMi06Mc07}
C. McNeile and C. Michael,
{\em Phys. Lett.} B 642 (2006) 244, hep-lat/0607032.
C. McNeile,
\newblock {L}attice {A}pproach to {L}ight {S}calars,
\newblock arXiv:0710.2470 [hep-lat]

\end{thebibliography}
\end{document}